\documentclass[reprint,aps,prb,superscriptaddress,twocolumn,longbibliography,floatfix]{revtex4-1}
\input{glyphtounicode}
\usepackage[T1]{fontenc}
\usepackage[utf8]{inputenc}
\usepackage{graphics,epsfig,amsfonts,amssymb,amsmath,color}
\usepackage{accents}
\usepackage{bbm}
\usepackage{color}
\usepackage{comment}
\usepackage{enumitem}
\usepackage{natbib}
\usepackage[colorlinks=true,linkcolor=black,citecolor=black,urlcolor=black,hypertexnames=false]{hyperref}
\usepackage{textcomp}
\usepackage{setspace}

\newcommand{\ba}{\begin{array}{rl}}
	\newcommand{\ea}{\end{array}}

\DeclareRobustCommand*{\citen}[1]{%
  \begingroup
    \romannumeral-`\x 
    \setcitestyle{numbers}%
    \cite{#1}%
  \endgroup   
}

\newcommand{\beginsupplement}{%
        \setcounter{section}{0}
        \renewcommand{\thesection}{\Alph{section}}%
        \setcounter{figure}{0}
        \renewcommand{\thefigure}{S\arabic{figure}}%
     }

\DeclareMathOperator{\Tr}{Tr}

\begin{document}



\title{A high-sensitivity charge sensor for silicon qubits above one kelvin}

\author{Jonathan Y. Huang} 
\email{yue.huang6@unsw.edu.au}
\affiliation{Centre for Quantum Computation \& Communication Technology, School of Electrical Engineering and Telecommunications, The University of New South Wales, Sydney 2052, Australia}
\author {Wee Han Lim}
\affiliation {Centre for Quantum Computation \& Communication Technology, School of Electrical Engineering and Telecommunications, The University of New South Wales, Sydney 2052, Australia}
\author {Ross C. C. Leon}
\affiliation {Centre for Quantum Computation \& Communication Technology, School of Electrical Engineering and Telecommunications, The University of New South Wales, Sydney 2052, Australia}
\author {Chih Hwan Yang}
\affiliation {Centre for Quantum Computation \& Communication Technology, School of Electrical Engineering and Telecommunications, The University of New South Wales, Sydney 2052, Australia}
\author {Fay E. Hudson}
\affiliation {Centre for Quantum Computation \& Communication Technology, School of Electrical Engineering and Telecommunications, The University of New South Wales, Sydney 2052, Australia}
\author {Christopher C. Escott}
\affiliation {Centre for Quantum Computation \& Communication Technology, School of Electrical Engineering and Telecommunications, The University of New South Wales, Sydney 2052, Australia}
\author {Andre Saraiva}
\email{a.saraiva@unsw.edu.au}
\affiliation {Centre for Quantum Computation \& Communication Technology, School of Electrical Engineering and Telecommunications, The University of New South Wales, Sydney 2052, Australia}
\author{Andrew S. Dzurak} 
\affiliation{Centre for Quantum Computation \& Communication Technology, School of Electrical Engineering and Telecommunications, The University of New South Wales, Sydney 2052, Australia}
\author {Arne Laucht}
\email{a.laucht@unsw.edu.au}
\affiliation {Centre for Quantum Computation \& Communication Technology, School of Electrical Engineering and Telecommunications, The University of New South Wales, Sydney 2052, Australia}

\date{\today}


\begin{abstract}
\noindent\textbf{ABSTRACT:}~Recent studies of silicon spin qubits at temperatures above 1\,K are encouraging demonstrations that the cooling requirements for solid-state quantum computing can be considerably relaxed. However, qubit readout mechanisms that rely on charge sensing with a single-island single-electron transistor (SISET) quickly lose sensitivity due to thermal broadening of the electron distribution in the reservoirs. Here we exploit the tunneling between two quantised states in a double-island SET (DISET) to demonstrate a charge sensor with an improvement in signal-to-noise by an order of magnitude compared to a standard SISET, and a single-shot charge readout fidelity above 99\,\% up to 8\,K at a bandwidth $>100$\,kHz. These improvements are consistent with our theoretical modelling of the temperature-dependent current transport for both types of SETs. With minor additional hardware overheads, these sensors can be integrated into existing qubit architectures for high fidelity charge readout at few-kelvin temperatures.~\\\\\textbf{KEYWORDS:}~Quantum computing, silicon, quantum dot, single-electron transistor, charge sensing, temperature
\end{abstract}

\maketitle


\section{Introduction}

Quantum computers 
may be realised by exploiting electron or nuclear spins as quantum bits (qubits) in silicon.\cite{Loss-1998,Kane1998,Morton2011,Zwanenburg2013,Veldhorst2017,Vandersypen2017} The spins of electrons in lithographically defined quantum dots have proven to be a prime candidate for a scalable quantum computer architecture by virtue of their long coherence time, high fidelity and compatibility with modern metal-oxide-semiconductor (MOS) technology.\cite{Ladd2018,gonzalezzalba2020} 
Most early demonstrations of this technology were performed in dilution refrigerators at temperatures around 100\,mK, where the cooling power of the cryostats is strongly limited and orders of magnitude smaller than at few-kelvin and beyond.\cite{Jazaeri2009,Hornibrook2015,Degenhardt2017}

Going to higher temperatures, the qubits may be operated in isolated mode, decoupled from the thermally broadened electron reservoir~\cite{Bayer2019,Yang2020} and read out using Pauli spin blockade.~\cite{Jones2018,Fogarty2018,Zhao2019,Yang2020,Seedhouse2020} Demonstrations of one- and two-qubit operation above 1\,K were recently performed with these techniques~\cite{Yang2020,Petit2020}. Both works report reduced readout visibility at high temperatures, ascribed to the decreased sensitivity of the employed SISETs. This is because the broadened electron energy distributions in the leads allow for current even when the dot chemical potential is not aligned between the reservoir chemical potentials. The reduction in signal also limits the readout bandwidth, which will be problematic for detecting and correcting qubit errors -- a problem that is amplified by the degraded spin lifetime and coherence time with increasing temperature. \cite{Petit2018,Yang2020}


In this letter we develop and test a temperature-resistant charge sensor based on a DISET, and compare it directly to a SISET obtained by disabling one of the tunneling barriers in the same device. We extract the amplitude and full-width-half-maximum (FWHM) of the Coulomb peaks of the DISET and the SISET from 1.7\,K to 12\,K and find good agreement with our theoretical model. Furthermore, we sense charge transitions in a  nearby quantum dot both in low-frequency AC lock-in\cite{Elzerman2004-2,Yang2012} and single-shot\cite{Gotz2008} measurements. We quantify the sensing performance for both schemes by calculating the signal-to-noise ratios (SNR) and error probabilities for single-shot sensing. The DISET offers significantly higher SNR and fault-tolerant charge readout\cite{Fowler2009,Fowler2012,Knill2005} up to 8\,K at the full nominal bandwidth of 200 kHz of our setup. Under the same conditions, the signal of the SISET is already below the noise amplitude. Taking into account the impact of temperature on spins~\cite{Yang2020} and measurement bandwidth, we determine that fault-tolerant readout fidelities ($>99.9$\,\%) can only be achieved with a SISET operated below 1.6\,K while the DISET can operate above 4.2\,K, conveniently achievable with a liquid ${}^4$He cryostat.


\section{Experimental Methods}\label{experimental_methods}

\begin{figure}
\centering
\includegraphics[width=3.1in]{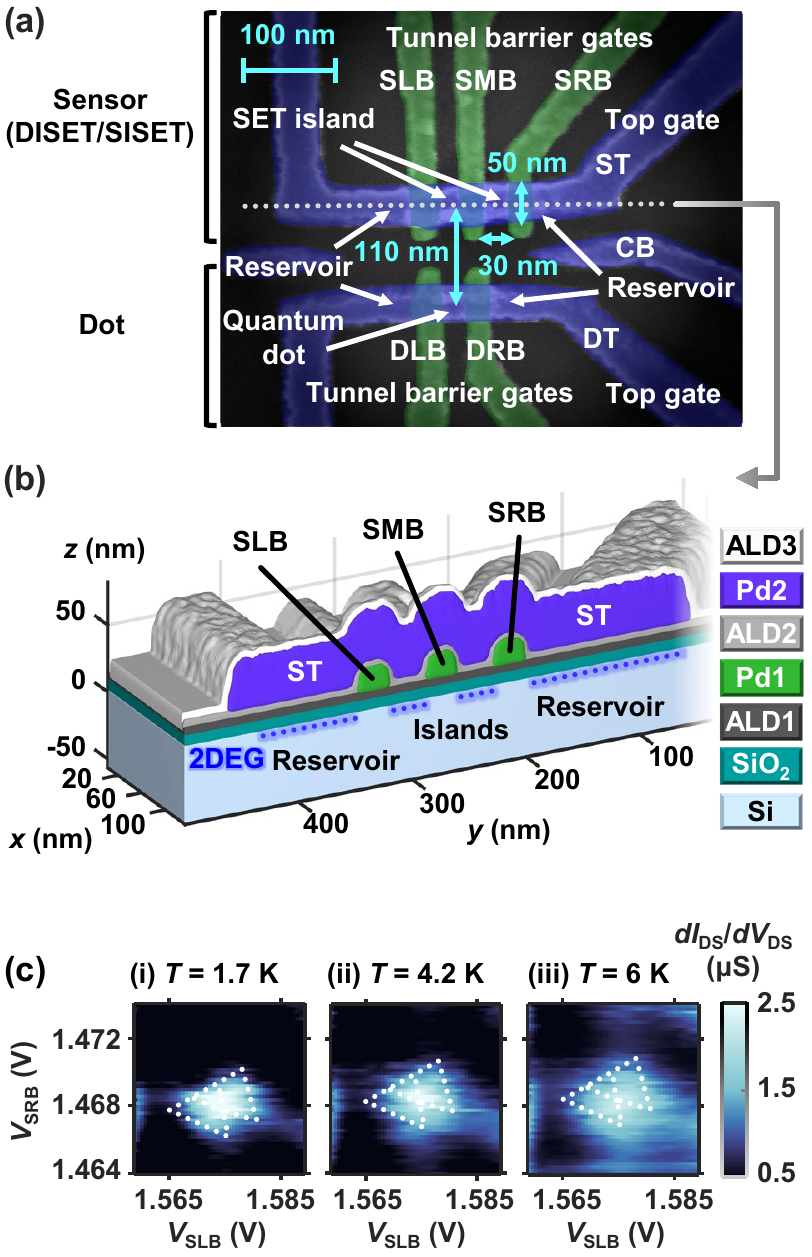}
\caption{Device overview and DISET transport spectroscopy. (a) False-coloured SEM image of a device from the same batch as the one measured. The bottom layer (green) consists of all barrier gates and the top layer (blue) consists of all top gates and the CB gate. In this work, the upper structure is configured as a sensor in either DISET or SISET mode, and the lower structure is configured as a single quantum dot. The lateral island/dot size is estimated to be 30\,nm\,$\times$\,50\,nm and the island-dot spacing is 110\,nm. With the CB gate not completely formed, the confinement of the islands and the dot relies mostly on the tunnel barrier gates and the lack of positive bias outside the top gate regions. (b) Simulated 3D structure of the DISET where the MOS structure can be seen. (c) In the single-electron tunneling regime with finite source-drain bias, bias triangle pairs are observed as a result of double-island transport. Shown here is the evolution of a typical triangle pair as the temperature increases.}
\label{fig1}
\end{figure}

Figure~\ref{fig1}(a) is a scanning electron micrograph (SEM) of a device nominally identical to the one measured in this work. It consists of a $^{\mathrm{nat}}$Si substrate, thermally grown SiO$_2$, and a two-layer Pd gate stack with atomic-layer-deposited Al$_x$O$_y$ (ALD) in between (details in Supporting Information~\ref{supp_A}). A top gate accumulates a two-dimensional electron gas (2DEG) under the Si/SiO${}_2$ interface, forming the electron reservoirs [see Figure~\ref{fig1}(b)]. The tunnel barrier gates then electrostatically define the sensor quantum dot\cite{Lim2009,Lim2009-2,Angus2007} as Figures~\ref{fig1}(a)-(b) indicate. If the sensor SRB is biased to form a potential barrier, it constricts the transport to single-electron tunneling and the sensor is configured as a DISET. However, if SRB is biased above threshold, the reservoir is extended and the sensor becomes a SISET using only the left island from the DISET. The dot in the lower structure in Figure~\ref{fig1}(a) acts as our device-under-test for the charge sensing measurements.

\section{Transport}\label{transport}

Transport characteristics are measured using an AC lock-in amplifier with a source-drain excitation of 0.2\,mV at 191\,Hz. Figure~\ref{fig1}(c) shows the differential conductance of the DISET as a function of the voltage on SLB and SRB. We apply a DC source-drain bias $V_{\mathrm{DS}}=(1\pm0.2)$\,mV, which creates a triangular region in voltage space with enhanced current (called bias triangles).\cite{Lim2009,Wiel2002} These regions occur near the triple points where states with different charge configurations are degenerate. The sensor top gate ST is biased at 2.92\,V, much higher than the turn-on voltage of 1.66\,V in this device, to enhance current. The middle barrier gate between the two sensor islands SMB is biased at 1.26\,V, in which case cotunneling is mostly suppressed. Figure~\ref{fig1}(c) shows a typical triangle pair in a well-tuned regime, of which a larger scan is shown in Supporting Information~\ref{supp_B}. The triangle pair broadens progressively as we increase the device temperature from 1.7\,K to 6\,K. The sides of the triangles correspond to the alignment of the Fermi level in the reservoirs to the energy level in the neighbouring island, whereas the base corresponds to the alignment of the energy levels of the two islands.\cite{Wiel2002} Comparing Figure~\ref{fig1}(c)(i)-(iii), we see more broadening on the triangle sides when the temperature increases, as expected. Additionally, cotunneling and background current appear more prominent with increasing temperature while the peak current reduces.

\begin{figure}[ht!]
\includegraphics[width=3.1in]{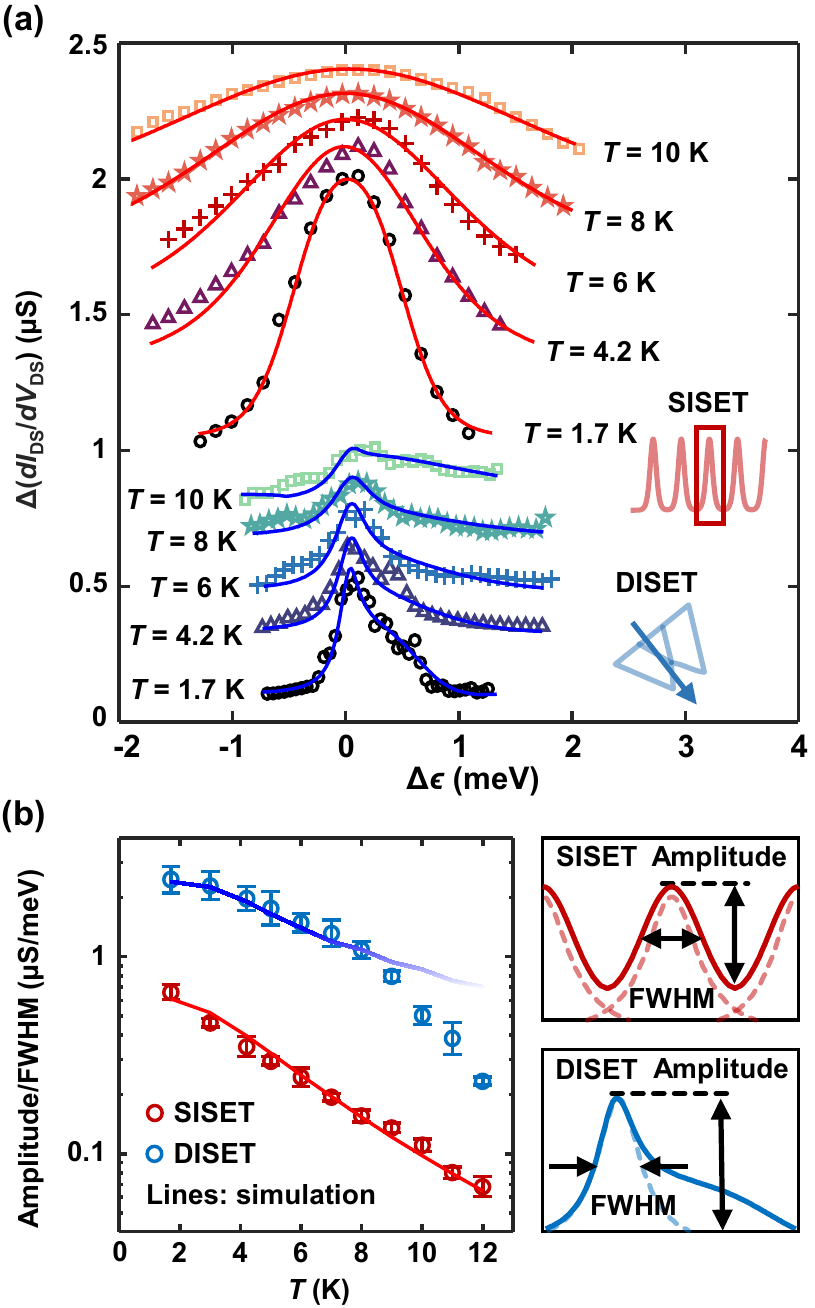}
\caption{Temperature dependence of transport characteristics. 
(a) Measured Coulomb peaks for the SISET and the DISET (markers) fitted with simulated curves (lines), at five temperatures from 1.7\,K to 10\,K (curves vertically offset for clarity). The differential conductance is plotted as a function of change in electrostatic potential $\Delta\epsilon$. Insets: The SISET’s peaks are obtained by sweeping ST, while the DISET’s peaks are obtained by sweeping SLB and SRB, which detunes the energy levels in the left and the right island. 
(b) Ratio of the peak amplitudes to their FWHM, which indicates the peaks’ steepness when $V_\mathrm{DS}$ is small. The experimental data (markers) are fitted with simulated curves (lines). The insets show the definition of amplitude and FWHM in the specific context of SISET (red) and DISET (blue).}
\label{fig2}
\end{figure}

In order to study the DISET sensitivity to electrostatic potential changes in the environment, we measure the differential conductance in response to detuning $\Delta\epsilon$, which is converted from gate voltage using relevant lever arms\cite{Lim2009,Fujisawa1998} (Supporting Information~\ref{supp_B}). The steepness of the peaks can be estimated using the amplitude divided by the full-width-half-maximum (FWHM), which quantifies the change in differential conductance per unit potential shift. Figure~\ref{fig2} shows a comparison between the SISET and the DISET, both biased for maximal slope to the best of our ability. We acquire the SISET peaks by sweeping ST and the DISET peaks by sweeping SLB and SRB in opposite directions, which is equivalent to a line cut through a triangle [see Figure~\ref{fig2}(a)-insets]. We see from Figure~\ref{fig2}(a) that throughout the temperature sweep, the DISET peak is evidently steeper at $\Delta\epsilon\approx0$ compared to $\Delta\epsilon>0$, and the broadening appears to be more evident as the temperature rises. We extract the FWHM from this dominant peak at $\Delta\epsilon\approx0$ and compare the extracted slope to that of the SISET in Figure~\ref{fig2}(b). At the same temperature, the DISET offers markedly larger slope than the SISET and this advantage increases until around 8\,K. At that point, the DISET degrades faster than the SISET, which we associate to the limit where the thermal broadening becomes larger than the Coulomb blockade quantisation window in the islands.

We develop a theoretical model for the transport through the quantised DISET levels to understand the experimental transport characteristics. The details of the model and analytical treatment of the equations are presented in Supporting Information~\ref{supp_C}. The resulting expression for the total DISET current is

\begin{equation}
    \label{DISET_current}
    I_{\mathrm{DS}} = e\frac{\Gamma_{02}\Gamma_{21}\Gamma_{10}-\Gamma_{01}\Gamma_{12}\Gamma_{20}+(\Gamma_{01}\Gamma_{20}-\Gamma_{10}\Gamma_{02})\Delta}{\Gamma_{\Sigma}},
\end{equation}
where $e$ is the unit charge,
\begin{equation}
    \label{Delta}
    \Delta = {t_{\mathrm{LR}}^2}\frac{\Gamma_{10}+\Gamma_{20}+\Gamma_{\mathrm{LR}}}{(\frac{\Gamma_{10}+\Gamma_{20}+\Gamma_{\mathrm{LR}}}{2})^2+(\frac{\mu_{\mathrm{L}}-\mu_{\mathrm{R}}}{h})^2}
\end{equation}
 and $\Gamma_{\Sigma}$ is the sum of all numerators. The transition rates $\Gamma$ are defined as follows:
 
 \begin{equation}
    \label{Gamma_01}
    \Gamma_{01} = f_{\mathrm{fD}}(\mu_{\mathrm{S}},T;\mu_{\mathrm{L}}) t_{\mathrm{S}}, \quad \Gamma_{10} = t_{\mathrm{S}} - \Gamma_{01},
\end{equation}
\begin{equation}
    \label{Gamma_20}
    \Gamma_{02} = f_{\mathrm{fD}}(\mu_{\mathrm{D}},T;\mu_{\mathrm{R}}) t_{\mathrm{D}}, \quad \Gamma_{20} = t_{\mathrm{D}} - \Gamma_{02},
\end{equation}
\begin{align}
    \label{Gamma_12_21}
    &\begin{cases}
    \Gamma_{12} = f_{\mathrm{fD}}(\mu_{\mathrm{L}},T;\mu_{\mathrm{R}})\Gamma_{\mathrm{LR}}\\ \Gamma_{21} = \Gamma_{\mathrm{LR}}-\Gamma_{12}\\
    \end{cases},
    \quad \mu_{\mathrm{L}} > \mu_{\mathrm{R}}\\
    &\begin{cases}
    \Gamma_{21} = f_{\mathrm{fD}}(\mu_{\mathrm{R}},T;\mu_{\mathrm{L}})\Gamma_{\mathrm{LR}}\\ \Gamma_{12} = \Gamma_{\mathrm{LR}}-\Gamma_{21}\\ 
    \end{cases},
    \quad \mu_{\mathrm{L}} \leq \mu_{\mathrm{R}}.
\end{align}

$f_{\mathrm{fD}}(\mu_{\mathrm{S}},T;\mu_{\mathrm{R}})$ and $f_{\mathrm{fD}}(\mu_{\mathrm{D}},T;\mu_{\mathrm{R}})$ describe the Fermi-Dirac distribution in the source and drain reservoirs, whose Fermi levels are $\mu_{\mathrm{S}}$ and $\mu_{\mathrm{D}}$ as set by the source-drain bias. The temperature is denoted by $T$, and $\mu_\mathrm{L}$ and $\mu_\mathrm{R}$ denote the energy levels in the left and right island respectively. The relevant rates are the reservoir-island tunnel rate $t_\mathrm{S}$, $t_\mathrm{D}$, inter-island tunnel rate $t_\mathrm{LR}$ and inter-island relaxation rate $\Gamma_\mathrm{LR}$.

\begin{figure*}[ht!]
\includegraphics[width=6.365in]{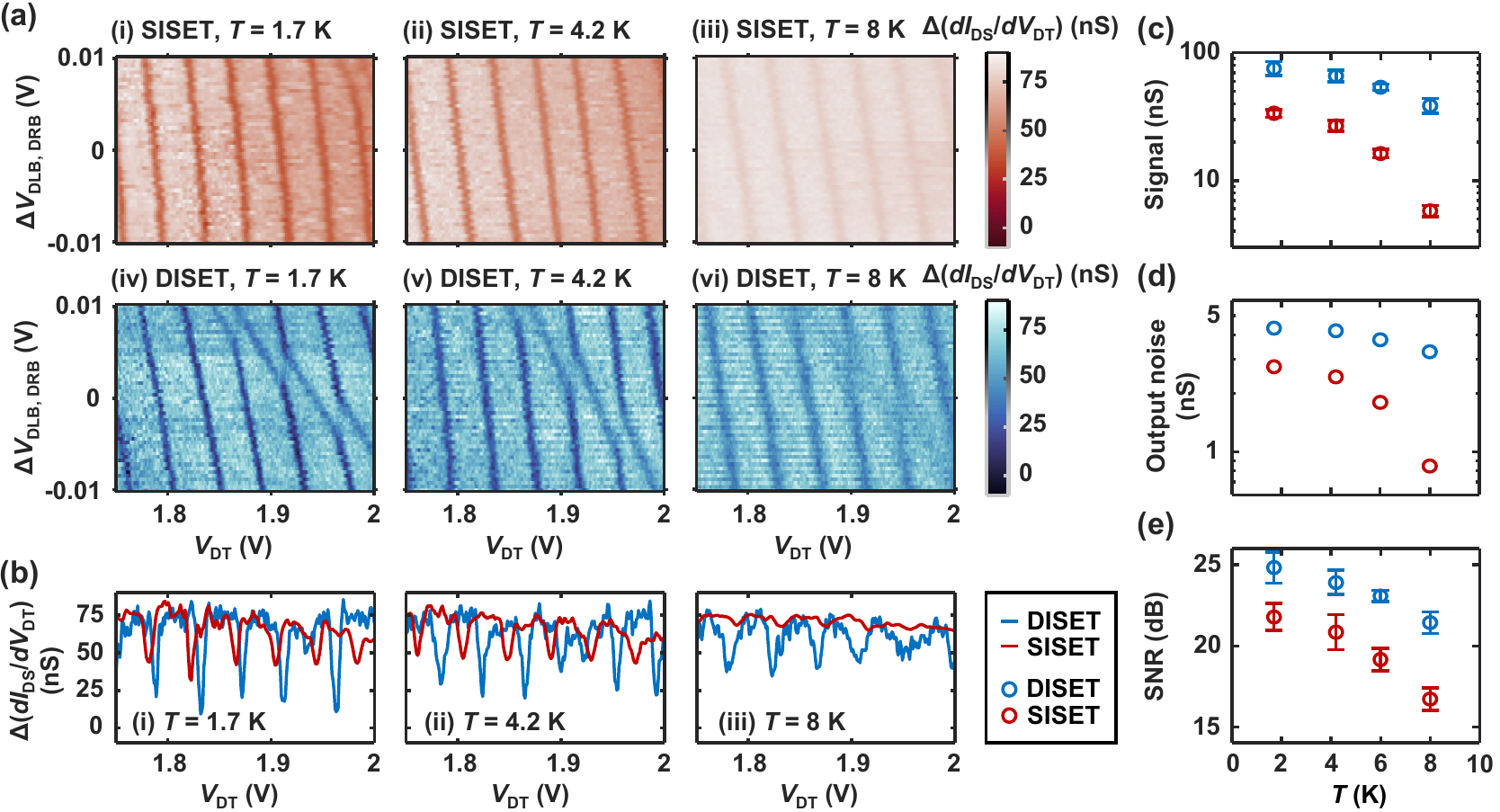}
\caption{Temperature dependence of lock-in charge sensing. 
(a) Signal from lock-in sensing of the single dot when the sensor is configured as (i) - (iii) SISET and (iv) - (vi) DISET at 1.7\,K, 4.2\,K, and 8\,K. Both sensing configurations are able to detect a number of charge transitions as $V_\mathrm{DT}$ is swept along the horizontal axis and $V_\mathrm{DLB}$, $V_\mathrm{DRB}$ are swept simultaneously along the vertical axis from 1.08 to 1.1\,V and from 1.19 to 1.21\,V, respectively. 
(b) We plot the 1D traces taken at $V_\mathrm{DLB} = 1.09$\,V and $V_\mathrm{DRB} = 1.2$\,V from (a). All traces are vertically aligned to visualise the difference in the signal amplitude. 
(c) - (e) Lock-in sensing signal, noise and signal-to-noise ratio (SNR) as a function of temperature. The average signal amplitude is calculated from the traces in (b). The noise level is extracted by taking the standard deviation in the flat parts of the traces. The SNR is then evaluated using the signal and noise level.}
\label{fig3}
\end{figure*}

Using this model, we tune the DISET and the SISET to their near-optimal sensing regime (details in Supporting Information~\ref{supp_D}). In Figure~\ref{fig2}, we bias the DISET at $V_\mathrm{SLB}=1.58$\,V, $V_\mathrm{SMB}=1.26$\,V and $V_\mathrm{SRB}=1.57$\,V. To operate the SISET, we apply $V_{\mathrm{SLB}}=1.54$\,V and $V_{\mathrm{SMB}}=1.26$\,V, while setting $V_{\mathrm{SRB}}=2$\,V to extend the reservoir into the DISET right island location. The peaks are further enhanced when a DC bias $V_\mathrm{DS}=(0.5\pm0.2)$\,mV is applied. We choose a Coulomb peak near $V_{\mathrm{ST}}=2.8$\,V to compare the SISET sensitivity to that of the DISET.

\begin{figure*}[ht!]
\includegraphics[width=6.4in]{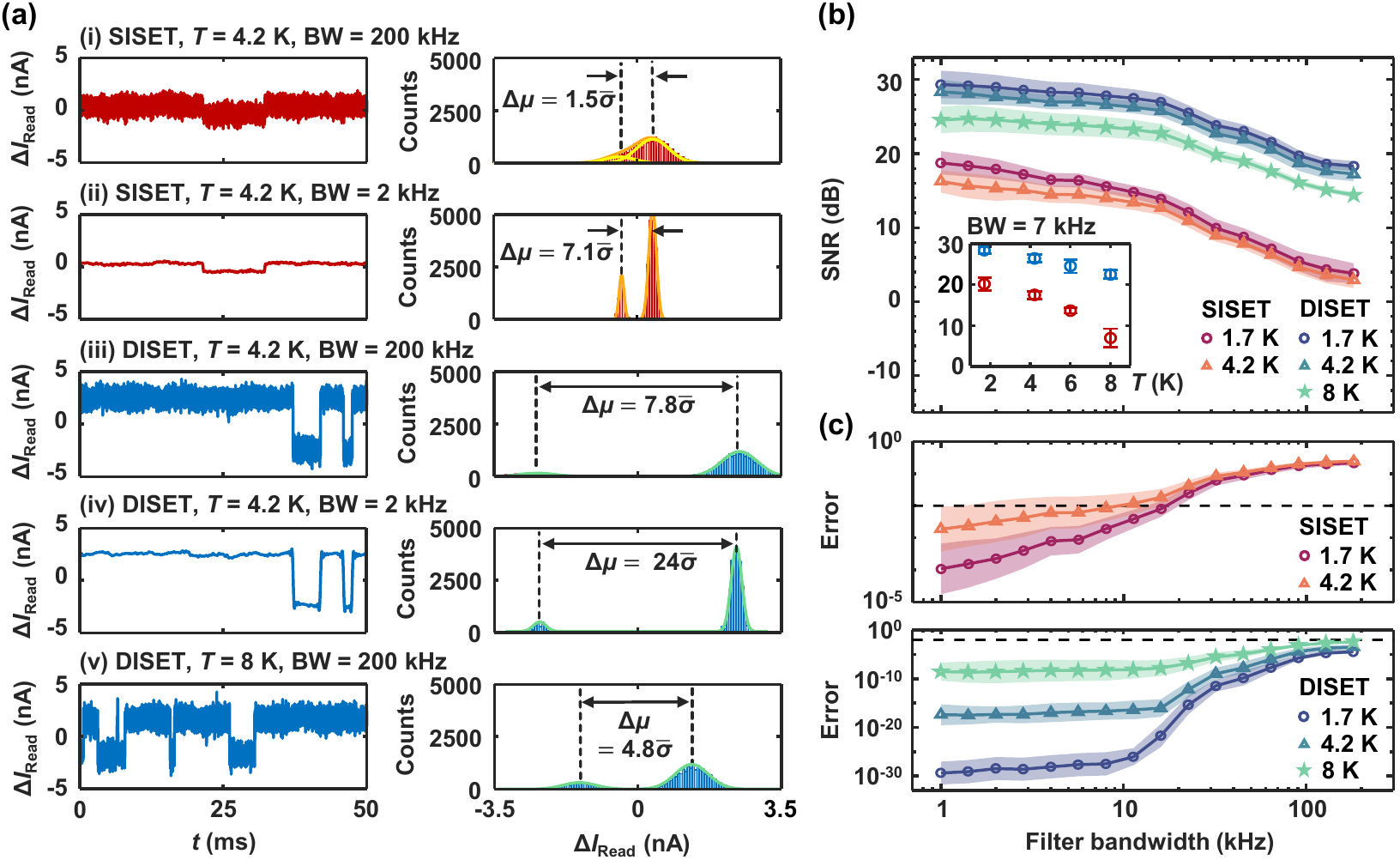}
\caption{Temperature dependence of single-shot charge readout. (a) Single-shot charge sensing signal when the sensor is configured as (i) - (ii) SISET and (iii) - (v) DISET. The measurement is taken at a nominal bandwidth of 200\,kHz and sample rate $f_\mathrm{s}=500$\,kHz. The traces are then numerically filtered to emulate various measurement bandwidths. The traces here show a comparison between a bandwidth of 200\,kHz and 2\,kHz. Histograms are plotted with a bin width of 60\,pA and fitted to Gaussian distribution, from which SNR and readout error can be extracted. 
(b) SNR as a function of bandwidth at 1.7\,K and 4.2\,K for the SISET and 1.7\,K, 4.2\,K and 8\,K for the DISET. Here the filter bandwidth ranges from 1\,kHz to 200\,kHz, and 30 traces are taken at each temperature. Inset: Temperature dependence measured separately at a bandwidth of 7\,kHz. The SNR here is slightly higher than the filtered data from the 200\,kHz measurement as the transimpedance amplifier is in a lower-bandwidth configuration. 
(c) Average readout error with logarithmic error bounds as a function of bandwidth, estimated from the fitting with Gaussian models.\cite{Yang2020,Medford2013} The dashed line marks the 1\,\% error threshold.}
\label{fig4}
\end{figure*}

The measured DISET peaks in Figure~\ref{fig2}(a) are fitted to Equation~\ref{DISET_current} without major disagreement (error $<25$\,\%), confirming the quantitative validity of the model. At low temperature, the error is dominated by current fluctuations due to charge noise. Starting from 6\,K, we see a different type of broadening around the resonant peak, which signifies additional physical processes. The steeper slope at $\Delta\epsilon<0$ stems from inter-island resonant tunneling.\cite{Stoof1996,Fujisawa1998} At $\Delta\epsilon>0$, inter-island relaxation current due to acoustic phonon emission also contributes, forming a shoulder that partially masks the resonant peak.\cite{Fujisawa1998,Wang2013} This relaxation current has negligible effects on the quality of the DISET current peaks. We also notice that the width of the shoulder deviates from $eV_\mathrm{DS}$ in some of the peaks, possibly because the sweep direction is not exactly perpendicular to the triangle base, or there is variation in $V_\mathrm{DS}$ and $\Gamma_\mathrm{LR}$ especially at high temperatures. In Figure~\ref{fig2}(b), after necessary conversions and assuming constant rates, we fit the data to Equation~\ref{DISET_current} with error $<5$\,\% up to 8\,K. We extract $t_\mathrm{LR}\approx20$\,GHz and the overall contribution from $t_\mathrm{S}$ and $t_\mathrm{D}$ $\sim40$\,GHz. Since $\Gamma_\mathrm{LR}$ has no influence on the DISET FWHM, we extract $\Gamma_\mathrm{LR}\approx6$\,GHz at 1.7\,K from the 1D peak in Figure~\ref{fig2}. We notice a deviation from 8\,K onward possibly attributable to orbital excitation\cite{Yang2012}, which is not considered in our model.

We also fit the SISET data to the expression for single-island transport current:\cite{ihn-3}

\begin{equation}
    \label{SISET_current}
    I_{\mathrm{DS}} = -e\frac{t_{\mathrm{S}} t_{\mathrm{D}}}{t_{\mathrm{S}} + t_{\mathrm{D}}}[f_{\mathrm{fD}}(\mu_{\mathrm{S}},T;\mu) - f_{\mathrm{fD}}(\mu_{\mathrm{D}},T;\mu)].
\end{equation}

We see no apparent deviation between the measured and simulated data. The fit errors in Figure~\ref{fig2}(a) and (b) are $<10$\,\%, mainly arising from potential fluctuations at lower temperatures (which are negligible at higher temperatures). The average contribution from $t_\mathrm{S}$ and $t_\mathrm{D}$ is $\sim42$\,GHz, close to that in the DISET.

Equations~\ref{DISET_current} and~\ref{SISET_current} suggest a predominantly exponential decay in the slope of the DISET and the SISET Coulomb peaks over temperature. The decay is notably slower in the DISET case. These characteristics are verified by the experimental data up to 8\,K. Moreover, the offset between the two curves is related to the tunnel rates and source-drain bias in both SETs, which depend strongly on device tuning.

\section{Charge Sensing}\label{charge_sensing}

We now employ the DISET and the SISET as charge sensors. To capture charge transitions in the quantum dot, we apply an AC excitation to DT with an amplitude of 2\,mV and a frequency of 127\,Hz, then demodulate the sensor current at this frequency to obtain the signal correlated to $V_{\mathrm{DT}}$ variation.\cite{Yang2012,Elzerman2004-2} We operate the sensor near the optimal regime identified in Section~\ref{transport} and tune it to the edge of a Coulomb peak where the highest slope is found. We apply dynamical compensation\cite{Yang2011} on ST, SLB and SRB when sensing with the DISET and on ST when sensing with the SISET. We further increase $V_\mathrm{DS}$ to 1.5\,mV in the DISET to enlarge the triangles without causing apparent variation to the sensitivity. We sweep DT between 1.75\,V and 2\,V while also varying DLB, DRB by $(\pm 0.01)$\,V around 1.09\,V and 1.2\,V.

Figure~\ref{fig3}(a) shows the resulting signal for the SISET (top row) and the DISET (bottom row) from 1.7\,K to 8\,K. Both SET configurations are stable during the sweeps and detect a number of transitions. In addition, we also observe the transition from a strongly-coupled parasitic dot, as indicated by the line with dissimilar slope and smaller amplitude in Figure~\ref{fig3}(a) (i) and (iv) - (vi).\cite{Yang2011} Figure~\ref{fig3}(b) shows the 1D traces along $V_\mathrm{DT}$ at around $V_\mathrm{DLB} = 1.09$\,V and $V_\mathrm{DRB} = 1.2$\,V. From the traces, we quantify the charge sensing performance by calculating the signal, noise and SNR, as shown in Figure~\ref{fig3}(c) - (e). We calculate the signal by averaging over the dips in the traces and the noise by taking the standard deviation of the flat region in between transitions. The SNR is then given by $20\log(\mathrm{signal}/\mathrm{noise})$. We see that the DISET signal is about twice that of the SISET at 1.7\,K and becomes almost an order of magnitude larger at 8\,K. The noise exhibits a similar temperature dependence, since the SET islands are sensitive to any electrostatic noise in the environment. It can also be inferred that the degradation in the SET sensitivity outpaces the linear increase in charge noise over temperature.\cite{Petit2018} Overall, since the DISET is more sensitive to both the charge transition and charge noise, the SNR is limited to a few dB higher than in the SISET case.

Lastly, we assess the charge sensitivity of SISET and DISET by performing single-shot charge readout of the nearby quantum dot.\cite{Gotz2008} We tune the sensors to their sensitive points and the dot to a charge transition point using lock-in sensing.\cite{Eenink2019} We then switch off the lock-in excitation and measure time traces at a nominal bandwidth of 200\,kHz and a sample rate $f_\mathrm{s}=500$\,kHz. To study the effect of lower measurement bandwidths, we numerically filter the raw traces with Butterworth low-pass filters of varying cutoff frequencies. To approximate the band-limiting effect in the experiment, the filters are designed to have $f_\mathrm{pass}-f_\mathrm{stop}=0.5 (f_\mathrm{s}-f_\mathrm{stop})$, where $f_\mathrm{s}$ is the sample rate, $f_\mathrm{pass}$ and $f_\mathrm{stop}$ are the upper edge of the pass band and the lower edge of the stop band. Figures~\ref{fig4}(a)(i) - (iv) show the vertically aligned current traces and the corresponding histograms for the SISET and the DISET at 4.2\,K at a bandwidth of 200\,kHz and 2\,kHz. Each histogram pair is fitted to a double Gaussian with separation $\Delta \mu$ and average standard deviation $\bar{\sigma}$. In Figure~\ref{fig4}(a)(v), we demonstrate the DISET working at 8\,K and a full effective bandwidth of 200\,kHz, where the SISET signal is completely indistinguishable from the noise. We see that a higher temperature produces a smaller signal ($\Delta\mu$), while a higher bandwidth yields larger noise ($\bar{\sigma}$).  Figure~\ref{fig4}(b) shows the SNR calculated using $20\log(\Delta\mu/\bar{\sigma})$, as a function of bandwidth at 1.7\,K, 4.2\,K for both SETs, and additionally at 8\,K for the DISET. We average over 30 traces at each temperature. The upper bound for the bandwidth in our experiment is 200\,kHz, which corresponds to the nominal amplifier bandwidth and a digitally-unfiltered signal, and the lower bound is 1\,kHz, below which significant damping of the signal occurs. At the same temperature, the DISET SNR is at least 10\,dB higher than that of the SISET and the difference slightly increases towards higher frequencies. The SNRs of both SETs experience a rapid decrease after $\sim20$\,kHz, which can be associated with the noise spectrum shown in Supporting Information~\ref{supp_E}.

We also perform a temperature sweep at a bandwidth of 7\,kHz and plot the temperature dependence in the inset of Figure~\ref{fig4}(b). Here we see a decay consistent to the one observed earlier.  However, in comparison to the lock-in measurements in Figure~\ref{fig3}(e), we note that at 1.7\,K both SETs begin with a higher SNR which then follows a faster decay. Moreover, a more significant improvement in SNR is observed for the full temperature range. This is because in the lock-in measurement, low-frequency charge noise dominates, whereas in the single-shot measurement, higher-frequency noise from the measurement setup dominates. The data implies that the DISET holds an even larger advantage over the SISET in the high measurement bandwidth regime.

We estimate the readout error from the double Gaussian fits to the histograms in Figure~\ref{fig4}(a).\cite{Yang2020,Medford2013} Plotted in Figure~\ref{fig4}(c), the error rate shows an agreement with both the temperature and bandwidth dependence of the SNR. We consider the fault-tolerant threshold, which specifies the maximum allowable error in quantum computation given a certain error correction protocol. If we set this threshold to be 1\,\% (or readout fidelity $>99$\,\%), the SISET must measure at a bandwidth below 20\,kHz at 1.7\,K, or below 2\,kHz at 4.2\,K based on the estimation. On the other hand, this readout fidelity is achieved by the DISET at all measurement conditions in our experiment. Assuming that its performance is maintained at even higher bandwidths, we infer that the DISET would enable fault-tolerance for qubit temperatures exceeding 4.2K.\cite{Yang2020,Petit2020}

\begin{figure}[ht!]
	\includegraphics[width=3.1in]{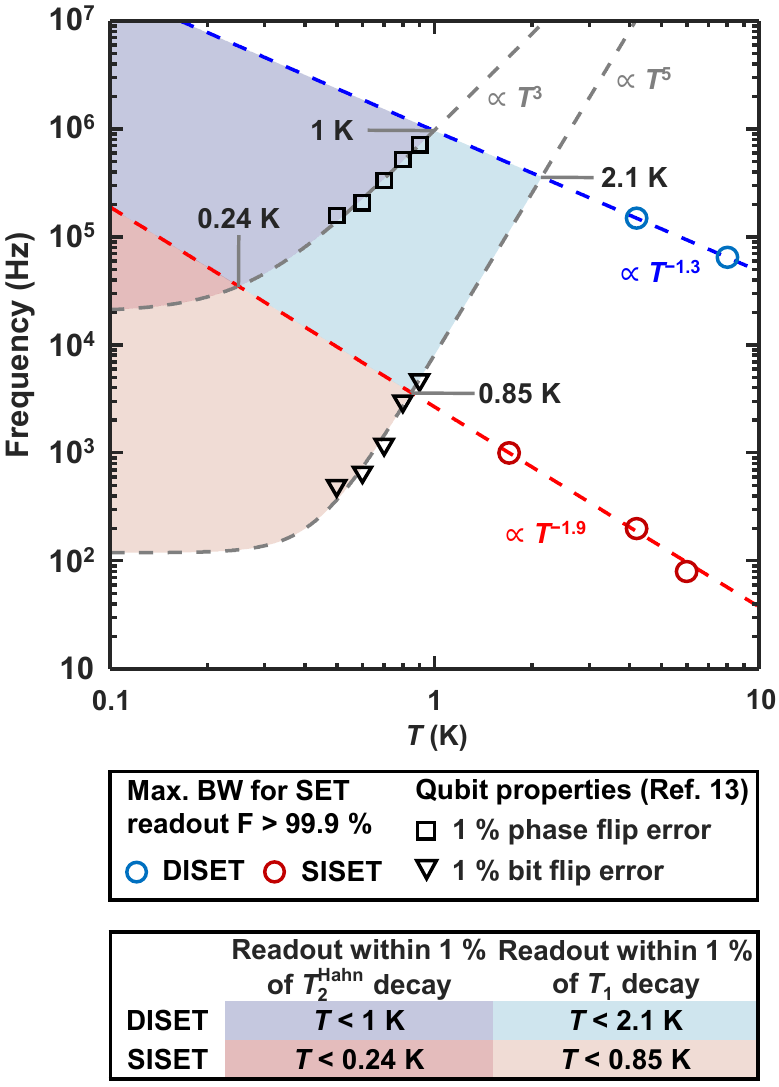}
	\caption{Temperature limit for qubit readout. The maximum bandwidth for readout fidelity $>99.9$\,\% in the DISET and the SISET are plotted and extrapolated with an inverse polynomial fit of $T^{-1.3}$ and $T^{-1.9}$ respectively, with the power of $T$ being a free fit parameter. Inverse of $T_1$ and $T_2^{\mathrm{Hahn}}$ from Reference~\citen{Yang2020} are also plotted with $T^5$ and $T^3$ fitting. The maximum temperatures are predicted from the crossings between $1/T_1$ and the maximum bandwidths.}
	\label{fig5}
\end{figure}

In Figure~\ref{fig5}, we numerically explore the limits in operation temperature. Now we set a readout error threshold of 0.1\,\% and extract the maximum allowable measurement bandwidth from the experiments in Figure~\ref{fig4}(c). This threshold value gives the maximum bandwidth for the DISET at 4.2\,K, 8\,K (blue circles) and that for the SISET at 1.7\,K, 4.2\,K and 6\,K (red circles). The other curves do not cross this threshold level. We extrapolate from the above data points using polynomial fitting, which shows a $T^{-1.3}$ dependence for the DISET (blue line) and a $T^{-1.9}$ dependence for the SISET (red line). We also plot the inverse of the time for reaching 99\,\% of $T_1$ and $T_2^{\rm Hahn}$ decay in a silicon qubit above 1\,K, calculated from the data in Reference~\citen{Yang2020}. The crossing between the curve for 99\,\% of $T_1$ and the maximum bandwidth for the sensor provides a crude indication of the maximum temperature for readout, since $T_1$ provides the ultimate limit in spin readout. Above this temperature, the faster decay of spins requires a higher readout bandwidth from the sensor, which will subsequently compromise on the readout fidelity due to the loss of sensitivity. In our device, such limit for the SISET and the DISET are 0.85\,K and 2.1\,K, respectively, which suggests that the DISET is competent in performing fault-tolerant readout above the base temperature in pumped $^4$He cryostats. 


\section{Conclusion and Outlook}
In conclusion, we have fabricated and measured a sensor that can be configured as either a SISET or a DISET to sense charge transitions in a nearby quantum dot at elevated temperatures. We observe the temperature dependence of Coulomb peaks when the SETs are well-tuned. We develop a theory for the DISET which accounts for both the inter-island resonant tunneling and relaxation, finding good agreement between the theory and experiment up to 8\,K. The DISET peaks offer appreciably larger slope at 1.7\,K and broaden at a $\sim50$\,\% slower rate when temperature is raised, as compared to the SISET in our operating regime. We apply both SETs in lock-in and single-shot charge sensing and demonstrate a considerably better signal with the DISET. This advantage becomes more pronounced with increasing temperature or higher measurement bandwidth. The DISET maintains a fidelity $>99$\,\% even at 8\,K, more than 100 times the temperature of the silicon qubits in dilution refrigerators\cite{Huang2019,Veldhorst2015,Veldhorst2014} and more than four times the temperature in recent hot qubit experiments.\cite{Yang2020,Petit2020}

Our single-shot measurement data suggest that a DISET, similar to the one we measured, can operate at above 4.2\,K, the temperature of liquid $^4$He, with a charge readout fidelity exceeding 99\,\%. In comparison, our SISET operation temperature is limited to 1.6\,K in order to achieve the same fidelity. This finding will potentially benefit qubit readout above 1\,K in the near future. Combined with the temperature-robustness of isolated mode operation,\cite{Yang2020,Petit2020} we expect to clear another challenge towards a silicon quantum computer that operates at few-kelvin, which is essential for the scalability of quantum processors.\\


\vspace{.2in}\noindent{\textbf{DATA AVAILABILITY}}

\noindent
The data supporting the findings in this study are available from the corresponding
authors upon reasonable request.

\vspace{.2in}\noindent{\textbf{ACKNOWLEDGEMENTS}}

\noindent
We thank Ensar Vahapoglu, Santiago Serrano and Tuomo Tanttu for assistance in the cryogenic measurement setup. We acknowledge support from the Australian Research Council (FL190100167 and CE170100012), the US Army Research Office (W911NF-17-1-0198)  and the NSW Node of the Australian National Fabrication Facility. The views and conclusions contained in this document are those of the authors and should not be interpreted as representing the official policies, either expressed or implied, of the
Army Research Office or the U.S. Government. The U.S. Government is authorized to reproduce and distribute reprints for Government purposes notwithstanding any copyright notation herein.

\vspace{.2in}


\noindent{\textbf{CONTRIBUTIONS}}

\noindent
JYH performed all measurements and all calculations. WHL and FH fabricated the devices. WHL, CHY and AL supervised all  measurements. WHL, RCCL, CHY, AS, AL and ASD participated in data interpretation and experiment planning. CHY, CE and AS supervised the numerical modelling. AS supervised the model development and analytical derivations. AS, AL and ASD designed the project and analysed the results. JYH, AS and AL wrote the manuscript with contributions from all authors.

\vspace{.2in}\noindent{\textbf{SUPPORTING INFORMATION}}

\noindent
The Supporting Information is available free of charge on the ACS Publications website at 
\href{https://doi.org/10.1021/acs.nanolett.1c01003}{https://doi.org/10.1021/acs.nanolett.1c01003}.~\\\\Device and measurement setup details, lever arms and tunability, DISET transport theory, transport characteristics and SET tuning, output noise spectrum.

\newpage
\onecolumngrid
\beginsupplement

\pagebreak
\widetext
\begin{center}
\large\textbf{Supporting Information:~A high-sensitivity charge sensor for silicon qubits above one kelvin}\par
\end{center}

\renewcommand{\thefigure}{S\arabic{figure}}
\def\theequation{S\arabic{equation}}

\section{Device and measurement setup details}\label{supp_A}

The device is fabricated on $^{\mathrm{nat}}$Si substrate with diffused $n^+$ source (S) and drain (D), defined by photolithography. A two-layer Pd gate stack is patterned by electron beam lithography (EBL) and deposited using physical vapour deposition (PVD), with atomic layer deposited (ALD) Al$_x$O$_y$ acting as insulator in between and around.\cite{Zhao2019,Brauns2018} Tunnel barrier gates for the sensor (SLB, SMB, SRB) and dot (DLB, DRB) are patterned together in the bottom layer (green); sensor and dot top gate (ST, DT) and confinement barrier gate (CB) are patterned in the top layer (blue). In all imaged devices, the CB gate is not fully formed, most likely due to the collapse of the polymethyl methacrylate (PMMA) resist that defines the geometry of the lithographically defined gates. We measure the device in an ICE Oxford variable temperature insert (VTI) that can operate from 1.7\,K to room temperature. DC biases on the gate electrodes are supplied by QDevil QDAC voltage generators. The AC lock-in measurements for transport and charge sensing are conducted using Stanford Research Systems SR830 lock-in amplifier. A FEMTO DLPCA-200 transimpedance amplifier is also used for amplification and filtering. In the single-shot measurements, time traces are recorded via a Pico Technology PicoScope 4824 digital oscilloscope, after amplification by the FEMTO DLPCA-200 amplifier and a Stanford Research Systems SR560 voltage amplifier.

\section{Lever Arms and Tunability}\label{supp_B}

\begin{figure}[h!]
\centering
\includegraphics[width=6.4in]{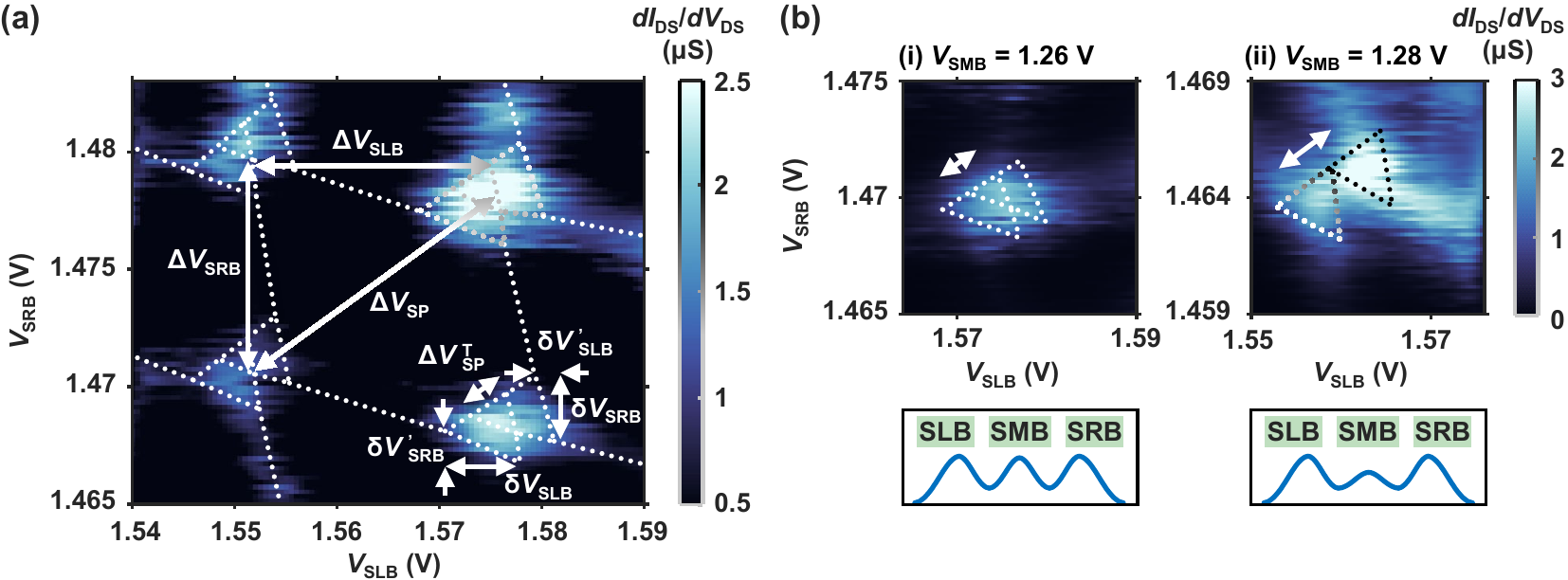}
\caption{Bias spectroscopy of the DISET. (a) Bias triangle pairs are observed as SLB and SRB are swept. Shown here are four neighbouring pairs, which form a well-defined charge stability map. Capacitances, lever arms and charging energies can be extracted from the dimensions of the charge stability map and the triangle pairs, which are traced out and labelled. (b) Splitting of a triangle pair occurs when $V_\mathrm{SMB}$ increases, creating a lower middle barrier between the two islands. All measurements were performed at 1.7\,K.}
\label{figS1}
\end{figure}

Lever arms are required for converting voltage detuning $\Delta V$ into energy detuning $\Delta\epsilon$, as well as to understand where the islands form in comparison to the gates. For the DISET, we first estimate the lever arm of SLB to the left island and that of SRB to the right island according to the formula $\alpha_{\mathrm{SLB}}^{\mathrm{L}}|e|\delta V_{\mathrm{SLB}} = \alpha_{\mathrm{SRB}}^{\mathrm{R}}|e|\delta V_{\mathrm{SRB}} = |e|V_{\mathrm{DS}}$.\cite{Lim2009,Lai2011,Wiel2002} We also consider the cross-coupling from SLB to the right island and from SRB to the left island, which give rise to $\alpha_{\mathrm{SLB}}^{\mathrm{R}}=\alpha_{\mathrm{SRB}}^{\mathrm{R}}{\delta V'_{\mathrm{SRB}}}/{\delta V_{\mathrm{SLB}}}$ and $\alpha_{\mathrm{SRB}}^{\mathrm{L}}=\alpha_{\mathrm{SLB}}^{\mathrm{L}}{\delta V'_{\mathrm{SLB}}}/{\delta V_{\mathrm{SRB}}}$. The voltage variations $\delta V_{\mathrm{SLB}}$, $\delta V_{\mathrm{SRB}}$, $\delta V'_{\mathrm{SLB}}$ and $\delta V'_{\mathrm{SRB}}$ are labelled in Figure~\ref{figS1}(a). Under the biasing condition in Figure~\ref{figS1}, we have $\alpha_{\mathrm{SLB}}^{\mathrm{L}}=0.25$, $\alpha_{\mathrm{SLB}}^{\mathrm{R}}=0.13$, $\alpha_{\mathrm{SRB}}^{\mathrm{L}}=0.23$ and $\alpha_{\mathrm{SRB}}^{\mathrm{R}}=0.38$. It appears that SRB has more control over both islands than SLB. When we increase $V_{\mathrm{SMB}}$ from 1.26\,V to 1.28\,V, $\alpha_{\mathrm{SLB}}^{\mathrm{L}}$ and $\alpha_{\mathrm{SRB}}^{\mathrm{R}}$ decrease to 0.13 and 0.30, suggesting that the islands are moving closer towards SMB.\cite{Eenink2019} However, $\alpha_{\mathrm{SLB}}^{\mathrm{R}}$ and $\alpha_{\mathrm{SRB}}^{\mathrm{L}}$ also decrease to 0.06 and 0.09, implying that the right island is less coupled to SLB and likewise for the left island. This is most likely because such cross-coupling is limited by the screening from the adjacent island, which becomes stronger as the islands come closer. Further increasing $V_{\mathrm{SMB}}$, we see the splitting of the triangles and the emergence of cotunneling lines [Figure~\ref{figS1}(b)(i)-(ii)], which signifies a transition from the weak-coupling regime to the strong-coupling regime. The double-island system is thus highly tunable.

\begin{figure}[h!]
\centering
\includegraphics[width=5.4in]{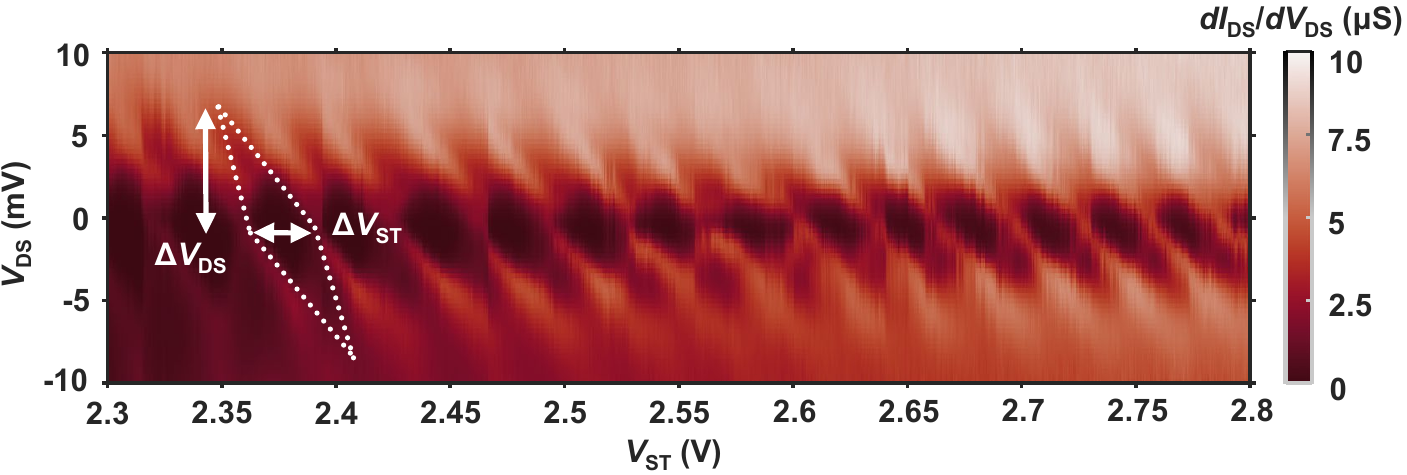}
\caption{Bias spectroscopy of the SISET. Coulomb diamonds are formed as ST is swept under different source-drain biases. Capacitances, lever arms and charging energies can be estimated from the dimension of adjacent diamond cells. This measurement was performed at 1.7\,K.}
\label{figS2}
\end{figure}

We also study an upper triangle pair at $V_\mathrm{SRB}\sim1.5$\,V but with similar $V_\mathrm{SLB}$ and $V_\mathrm{SMB}$. In this case, $\alpha_{\mathrm{SLB}}^{\mathrm{L}}=0.11$, $\alpha_{\mathrm{SLB}}^{\mathrm{R}}=0.07$, $\alpha_{\mathrm{SRB}}^{\mathrm{L}}=0.10$ and $\alpha_{\mathrm{SRB}}^{\mathrm{R}}=0.31$. The SRB lever arm $\alpha_{\mathrm{SRB}}^{\mathrm{R}}$ here is smaller than that with lower $V_\mathrm{SRB}$, which is understood to be caused by an increased occupancy. While the gate capacitance $C_\mathrm{{SRB}}^{\mathrm{R}}$ remains approximately unchanged, the total island capacitance $C_{\Sigma}^{\mathrm{R}}$ increases, therefore a smaller lever arm results according to $ {\alpha_{\mathrm{SRB}}^{\mathrm{R}}} = {C_\mathrm{{SRB}}^{\mathrm{R}}}/C_{\Sigma}^{\mathrm{R}}$.\cite{Wiel2002}

We notice that in all cases, $\alpha_\mathrm{SRB}^\mathrm{R}$ is almost twice as large as $\alpha_\mathrm{SLB}^\mathrm{L}$, indicating a much stronger coupling between SRB and the right island compared to that between SLB and the left island. For this reason, we see notably more Coulomb peaks when sweeping along the $V_\mathrm{SRB}$ direction in the window between pinch-off and saturation. We thus operate primarily around different $V_\mathrm{SRB}$ to observe the effect of reservoir-island tunnel rates.

Lastly, the lever arm of ST to the single island $\alpha_{\mathrm{ST}}$ is estimated using the formula $\alpha_{\mathrm{ST}}=\Delta V_{\mathrm{DS}}/\Delta V_{\mathrm{ST}}$,\cite{Lim2009-2,Angus2007} where $\Delta V_{\mathrm{DS}}$ and $\Delta V_{\mathrm{ST}}$ are labelled in Figure~\ref{figS2}. When ST is increased from $\sim2.5$\,V to $\sim2.8$\,V, $\alpha_\mathrm{ST}$ decreases from 0.18 to 0.14, showing a similar trend to $ {\alpha_{\mathrm{SLB}}^{\mathrm{L}}}$ and $ {\alpha_{\mathrm{SRB}}^{\mathrm{R}}}$ in the DISET.

\section{DISET Transport Theory}\label{supp_C}

In this section, we derive the expression for the total DISET current, as shown in Equation~\ref{DISET_current} in the main text. We describe the DISET with the Hubbard Model\cite{Hubbard1963,Sarma2011,Wang2011} and follow a Master Equation\cite{Lindblad1976,Gorini1976} approach. Previous studies have calculated either the inter-island/dot resonant tunneling current\cite{Stoof1996,Wiel2002,Fujisawa1998,Gurvitz1996,Gurvitz1996b} or the relaxation current\cite{Touati2014,Zhang2010,Boubaker2009,Brennerthesis} alone. Here we develop a complete model that accounts for both the elastic transport and the inelastic transport.

We consider the DISET as an open quantum system, in which electrons occupying discrete energy states in the islands also interact with the environment through the reservoirs. This allows us to study the electron occupation by solving the Lindblad master equation\cite{Lindblad1976,Gorini1976} for density matrix $\hat{\rho}$. We then calculate the steady state current from the electron flow rate, which can be derived from $\hat{\rho}$.

The Lindblad equation we use takes the form
\begin{equation}
    \label{lindblad_equation_S}
    \frac{d\hat{\rho}}{dt} = -i[\hat{H},\hat{\rho}] + \hat{\hat{\mathcal{L}}}(\hat{\rho}), 
\end{equation}
\vspace{0.2in}
where
\begin{equation}
    \label{dissipator_S}
     \hat{\hat{\mathcal{L}}}(\hat{\rho})=\sum_{k}\gamma_k(\hat{L_k}\hat{\rho} \hat{L_k}^\dagger - \frac{1}{2}\{ \hat{L_k}^\dagger \hat{L_k},\hat{\rho}\}).
\end{equation}

The Hamiltonian $\hat{H}$ is Hermitian and $\sqrt{\gamma_k}\hat{L_k}$ are jump operators. The first term in Equation~\ref{lindblad_equation_S} originates from the von Neumann equation and describes the elastic processes, such as the inter-island resonant tunneling in our case. The second term, known as the Lindbladian dissipator, describes inelastic or dissipative processes, such as the reservoir-island transport or the inter-island relaxation.

We describe our coupled double-island system using the Hubbard model Hamiltonian\cite{Hubbard1963,Sarma2011,Wang2011}
\begin{equation}
    \label{Hamiltonian_S}
    \hat{H} = 
    \begin{pmatrix}
    0 & 0 & 0\\
    0 & \mu_{\mathrm{L}} & t_{\mathrm{LR}}\\
    0 & t_{\mathrm{LR}} & \mu_{\mathrm{R}}\\
    \end{pmatrix},
\end{equation}
which is written in the basis $\{|0\rangle, |1\rangle$, $|2\rangle\}$ describing the state of no additional electrons, one additional electron in the left island and one additional electron in the right island (with respect to a certain number of electrons in either dots, depending on the particular bias triangle that we are operating on). The diagonal elements represent the total energy required for each state, while the off-diagonal elements represent the elastic tunnel coupling. For simplicity we set our zero energy to be aligned with the dot in the $|0\rangle$ state. The energies $\mu_{\mathrm{L}}$ and $\mu_{\mathrm{R}}$ comprise potential offsets, background charge, local charge repulsion and inter-island charge repulsion. Elastic coherent tunnel coupling is considered only between $|1\rangle$ and $|2\rangle$ -- all transitions involving the reservoirs are considered to be elastic but any coherence quickly decays. This means that the zero row and column in the Hamiltonian describe a state $|0\rangle$ that only couples to the other states mediated by the Lindbladian dissipator.

The Lindbladian dissipator reads
\begin{equation}
    \label{Lindbladian_S}
    \hat{L}(\hat{\rho}) = \hat{D}_{01} + \hat{D}_{12} + \hat{D}_{\mathrm{20}},
\end{equation}
in which
\begin{equation}
    \label{D_terms_S}
    \begin{cases}
    \hat{D}_{01} = \Gamma_{01} (\hat{\sigma}_{10} \hat{\rho}  \hat{\sigma}_{01} - \frac{1}{2}\{\hat{\sigma}_{10}  \hat{\sigma}_{01}, \hat{\rho}\}) + \Gamma_{10} (\hat{\sigma}_{01} \hat{\rho} \hat{\sigma}_{10} - \frac{1}{2}\{\hat{\sigma}_{01}  \hat{\sigma}_{10} ,\hat{\rho}\})\\
    \hat{D}_{12} = \Gamma_{12} (\hat{\sigma}_{21} \hat{\rho}  \hat{\sigma}_{12} - \frac{1}{2}\{\hat{\sigma}_{21}  \hat{\sigma}_{12}, \hat{\rho}\}) + \Gamma_{21} (\hat{\sigma}_{12} \hat{\rho} \hat{\sigma}_{21} - \frac{1}{2}\{\hat{\sigma}_{12}  \hat{\sigma}_{21} ,\hat{\rho}\})\\
    \hat{D}_{20} = \Gamma_{20} (\hat{\sigma}_{02} \hat{\rho}  \hat{\sigma}_{20} - \frac{1}{2}\{\hat{\sigma}_{02}  \hat{\sigma}_{20} ,\hat{\rho}\}) + \Gamma_{02} (\hat{\sigma}_{20} \hat{\rho} \hat{\sigma}_{02} - \frac{1}{2}\{\hat{\sigma}_{20}  \hat{\sigma}_{02} ,\hat{\rho}\})
    \end{cases}.
\end{equation}
The operators $\hat{\sigma}$ describe the inelastic transitions between states at rates $\Gamma$, which are determined as follows:
\begin{equation}
    \label{Gamma_01_S}
    \Gamma_{01} = f_{\mathrm{fD}}(\mu_{\mathrm{S}},T;\mu_{\mathrm{L}}) t_{\mathrm{S}}, \quad \Gamma_{10} = t_{\mathrm{S}} - \Gamma_{01},
\end{equation}
\begin{equation}
    \label{Gamma_20_S}
    \Gamma_{02} = f_{\mathrm{fD}}(\mu_{\mathrm{D}},T;\mu_{\mathrm{R}}) t_{\mathrm{D}}, \quad \Gamma_{20} = t_{\mathrm{D}} - \Gamma_{02},
\end{equation}
\begin{align}
    \label{Gamma_12_21_S}
    &\begin{cases}
    \Gamma_{12} = f_{\mathrm{fD}}(\mu_{\mathrm{L}},T;\mu_{\mathrm{R}})\Gamma_{\mathrm{LR}}\\ \Gamma_{21} = \Gamma_{\mathrm{LR}}-\Gamma_{12}\\
    \end{cases},
    \quad \mu_{\mathrm{L}} > \mu_{\mathrm{R}}\\
    &\begin{cases}
    \Gamma_{21} = f_{\mathrm{fD}}(\mu_{\mathrm{R}},T;\mu_{\mathrm{L}})\Gamma_{\mathrm{LR}}\\ \Gamma_{12} = \Gamma_{\mathrm{LR}}-\Gamma_{21}\\ 
    \end{cases},
    \quad \mu_{\mathrm{L}} \leq \mu_{\mathrm{R}}.
\end{align}
Here $t_{\mathrm{S}}$, $t_{\mathrm{LR}}$ and $t_{\mathrm{D}}$ are the tunnel rates between the source and the left island, between the left and the right island and between the right island and the drain. The inter-island relaxation rate is $\Gamma_{\mathrm{LR}}$, and $f_{\mathrm{fD}}$ is the Fermi-Dirac distribution in the reservoir, expressed as
\begin{equation}
    \begin{cases}
        f_{\mathrm{fD}}(\mu_{\mathrm{S}},T;\mu_{\mathrm{L}}) = \frac{1}{e^{\frac{\mu_{\mathrm{L}}-\mu_{\mathrm{S}}}{kT}}+1}\\
        f_{\mathrm{fD}}(\mu_{\mathrm{D}},T;\mu_{\mathrm{R}}) = \frac{1}{e^{\frac{\mu_{\mathrm{R}}-\mu_{\mathrm{D}}}{kT}}+1}
    \end{cases},
\end{equation}
where $T$ is the electron temperature and $\mu_{\mathrm{S}}$, $\mu_{\mathrm{D}}$ are the chemical potentials at the source and drain.

We solve Equation~\ref{lindblad_equation_S} by first vectorising $\hat{\rho}$ and rewriting the Lindblad equation (Equation~\ref{Lindbladian_S}) as $\dot{\rho}=(\mathcal{H}+\mathcal{L})\rho$

\begin{equation}
    \label{L_Mat_S}
    \mathcal{L} = 
    \begin{pmatrix}
    -\Gamma_{01}-\Gamma_{02} & 0 & 0 & 0 & \Gamma_{10} & 0 & 0 & 0 & \Gamma_{20}\\
    0 & L_{22} & 0 & 0 & 0 & 0 & 0 & 0 & 0\\
    0 & 0 & L_{33} & 0 & 0 & 0 & 0 & 0 & 0\\
    0 & 0 & 0 & L_{44} & 0 & 0 & 0 & 0 & 0\\
    \Gamma_{01} & 0 & 0 & 0 & -\Gamma_{10}-\Gamma_{12} & 0 & 0 & 0 & \Gamma_{21}\\
    0 & 0 & 0 & 0 & 0 & L_{66} & 0 & 0 & 0\\
    0 & 0 & 0 & 0 & 0 & 0 & L_{77} & 0 & 0\\
    0 & 0 & 0 & 0 & 0 & 0 & 0 & L_{88} & 0\\
    \Gamma_{02} & 0 & 0 & 0 & \Gamma_{12} & 0 & 0 & 0 & -\Gamma_{20}-\Gamma_{21}\\
    \end{pmatrix},
\end{equation}

\begin{equation}
    \label{Diagonals_S}
    \begin{cases}
    L_{22} = L_{44} = -\frac{\Gamma_{01} + \Gamma_{02} + \Gamma_{10} + \Gamma_{12}}{2} \\
    L_{33} = L_{77} = -\frac{\Gamma_{01} + \Gamma_{02} + \Gamma_{20} + \Gamma_{21}}{2} \\
    L_{66} = L_{88} = -\frac{\Gamma_{10} + \Gamma_{12} + \Gamma_{20} + \Gamma_{21}}{2}
    \end{cases}.
\end{equation}
Likewise, the von Neumann term (Equation~\ref{Hamiltonian_S}) can be reshaped as

\begin{equation}
    \label{H_Mat_S}
    \mathcal{H} = 
    \begin{pmatrix}
    0 & 0 & 0 & 0 & 0 & 0 & 0 & 0 & 0\\
    0 & -\frac{\mu_Li}{h} & -\frac{t_{LR}i}{h} & 0 & 0 & 0 & 0 & 0 & 0\\
    0 & -\frac{t_{LR}i}{h} & -\frac{\mu_Ri}{h} & 0 & 0 & 0 & 0 & 0 & 0\\
    0 & 0 & 0 & \frac{\mu_Li}{h} & 0 & 0 & \frac{t_{LR}i}{h} & 0 & 0\\
    0 & 0 & 0 & 0 & 0 & -\frac{t_{LR}i}{h} & 0 & \frac{t_{LR}i}{h} & 0\\
    0 & 0 & 0 & 0 & -\frac{t_{LR}i}{h} & \frac{(\mu_L-\mu_R)i}{h} & 0 & 0 & \frac{t_{LR}i}{h}\\
    0 & 0 & 0 & \frac{t_{LR}i}{h} & 0 & 0 & \frac{\mu_Ri}{h} & 0 & 0\\
    0 & 0 & 0 & 0 & \frac{t_{LR}i}{h} & 0 & 0 & -\frac{(\mu_L-\mu_R)i}{h} & -\frac{t_{LR}i}{h}\\
    0 & 0 & 0 & 0 & 0 & \frac{t_{LR}i}{h} & 0 & -\frac{t_{LR}i}{h} & 0\\
    \end{pmatrix}.
\end{equation}
Substitution of Equations~\ref{L_Mat_S} and \ref{H_Mat_S} into Equation~\ref{lindblad_equation_S} with the steady state condition $d\hat{\rho}/dt=0$ will result in a set of nine linear equations. We reduce the system of equations to those containing populations:
\begin{equation}
    \label{H_eq_mat_S}
    \vec{0} = 
    \begin{pmatrix}
    -\Gamma_{01}-\Gamma_{02} & \Gamma_{10} & 0 & 0 & \Gamma_{20}\\
    \Gamma_{01} & -\Gamma_{10}-\Gamma_{12} & -\frac{t_{LR}i}{h} & \frac{t_{LR}i}{h} & \Gamma_{21}\\
    0 & -\frac{t_{LR}i}{h} &\frac{(\mu_L-\mu_R)i}{h}- \frac{\Gamma_{10}+\Gamma_{12}+\Gamma_{20} +\Gamma_{21}}{2} & 0 & \frac{t_{LR}i}{h}\\
    0 & \frac{t_{LR}i}{h} & 0 & -\frac{(\mu_L-\mu_R)i}{h}- \frac{\Gamma_{10}+\Gamma_{12}+\Gamma_{20} +\Gamma_{21}}{2} & -\frac{t_{LR}i}{h}\\
    \Gamma_{02} & \Gamma_{12} & \frac{t_{LR}i}{h} & -\frac{t_{LR}i}{h} &  -\Gamma_{20}-\Gamma_{21}\\
    \end{pmatrix}
    \begin{pmatrix}
    \rho_{00}\\
    \rho_{11}\\
    \rho_{21}\\
    \rho_{12}\\
    \rho_{22}\\
    \end{pmatrix}.
\end{equation}
The solution to the populations in Equation~\ref{H_eq_mat_S} is given by
\begin{equation}
    \label{populations_S}
    \begin{cases}
    \rho_{00} = \frac{\Gamma_{21}\Gamma_{10}+\Gamma_{10}\Gamma_{20}+\Gamma_{12}\Gamma_{20}-\Gamma_{10}\Delta-\Gamma_{20}\Delta}{\Gamma_{\Sigma}}\\
    \rho_{11} =
    \frac{\Gamma_{20}\Gamma_{01}+\Gamma_{01}\Gamma_{21}+\Gamma_{02}\Gamma_{21}-\Gamma_{01}\Delta-\Gamma_{02}\Delta}{\Gamma_{\Sigma}}\\
    \rho_{22} =
    \frac{\Gamma_{01}\Gamma_{12}+\Gamma_{02}\Gamma_{12}+\Gamma_{02}\Gamma_{10}-\Gamma_{01}\Delta-\Gamma_{02}\Delta}{\Gamma_{\Sigma}}
    \end{cases},
\end{equation}
where
\begin{equation}
    \label{Delta_S}
    \Delta = {t_{\mathrm{LR}}^2}\frac{\Gamma_{10}+\Gamma_{20}+\Gamma_{\mathrm{LR}}}{(\frac{\Gamma_{10}+\Gamma_{20}+\Gamma_{\mathrm{LR}}}{2})^2+(\frac{\mu_{\mathrm{L}}-\mu_{\mathrm{R}}}{h})^2},
\end{equation}
 and $\Gamma_{\Sigma}$ is the sum of all numerators. Notice that the populations and currents only depend on the inter-dot elastic tunneling $t_{\mathrm{LR}}^2$ through $\Delta$, such that it completely determines how the tunnel barrier control will impact the performance of the DISET. Comparing this term to the derivations in Reference~\onlinecite{Brennerthesis}, our model gives special attention to this elastic tunneling, which is an important ingredient for the added robustness of our device against elevated temperatures.

The steady state current through the system can be calculated using the current operator $\hat{J}$:\cite{Stoof1996}
\begin{equation}
    \label{I_avg_S}
    \langle I \rangle = e\Tr(\hat{\rho} \hat{J}),
\end{equation}
where $\hat{J}$ is a sum over $\hat{J}_{m\rightarrow n}$, the probability current between state $|m\rangle$ and $|n\rangle$: \cite{Hovhannisyan2019}
\begin{equation}
    \label{Jmn}
    \hat{J}_{m\rightarrow n} = \frac{1}{2}\sum_k \gamma_k [\{\hat{\rho}_m,\hat{L}_k^\dagger \hat{\rho}_n \hat{L}_k\} - \{\hat{\rho}_n,\hat{L}_k^\dagger \hat{\rho}_m \hat{L}_k\}].
\end{equation}
Here $\hat{\rho}_m$ and $\hat{\rho}_n$ represent the basis states $| m \rangle \langle m | $ and $| n \rangle \langle n | $. This current originates from the equilibrium electronic state transitions and the expression is a statistical equivalent of the formula $I=dq/dt$. We also note that for our DISET with series-connected islands and reservoirs, the current continuity condition makes necessary that the transport current between any two states be equal in our three-state model. The above considerations lead us to substitute Equation~\ref{Jmn} into Equation~\ref{I_avg_S}, with $\hat{J}_{m\rightarrow n}$ trivially chosen as $\hat{J}_{0\rightarrow 1}$. We arrive at
\begin{equation}
    \label{I_01_S}
    I_{\mathrm{DS}}=\langle I_{0\rightarrow 1} \rangle = e(\rho_{00}\Gamma_{\mathrm{01}}-\rho_{11}\Gamma_{\mathrm{10}}).
\end{equation}
Using the result for $\rho_{00}$ and $\rho_{11}$ from Equation~\ref{populations_S}, we end up with 
\begin{equation}
    \label{DISET_current_S}
    I_{\mathrm{DS}} = -e\frac{\Gamma_{01}\Gamma_{12}\Gamma_{20}-\Gamma_{02}\Gamma_{21}\Gamma_{10}+(\Gamma_{01}\Gamma_{20}-\Gamma_{10}\Gamma_{02})\Delta}{\Gamma_{\Sigma}},
\end{equation}
where $\Delta$ and $\Gamma$ have been defined in Equation~\ref{Delta_S} and Equations~\ref{Gamma_01_S}-\ref{Gamma_12_21_S}.

This is the minimum model that can describe a DISET. Hence Equation~\ref{DISET_current_S} accounts for a single bias triangle. Nevertheless, this model combines resonant\cite{Stoof1996} and relaxation current\cite{Brennerthesis} as separately presented in previous literature. If we set $\Gamma_{\mathrm{LR}}=0$ and $T=0$, we recover the expression for a purely resonant process,
\begin{equation}
    \label{Stoof_formula_S}
    I_{\mathrm{DS}} = -e\frac{t_{\mathrm{D}}{t_{\mathrm{LR}}}^2}{(\frac{\mu_{\mathrm{L}}-\mu_{\mathrm{R}}}{h})^2+\frac{t_{\mathrm{D}}^2}{4}+{t_{\mathrm{LR}}}^2(2+\frac{t_{\mathrm{D}}}{t_{\mathrm{S}}})},
\end{equation}
as derived in Reference \citen{Stoof1996}. This is a Lorentzian function and is characteristic of a resonant peak in an SET without any dissipation. On the other hand, if we set $t_{\mathrm{LR}}=0$, $\Delta$ vanishes which leads to an expression of purely dissipative current:
\begin{equation}
    \label{Brenner_formula_S}
    I_{\mathrm{DS}} = -e\frac{\Gamma_{01}\Gamma_{12}\Gamma_{20}-\Gamma_{02}\Gamma_{21}\Gamma_{10}}{\Gamma_{\Sigma}}
\end{equation}
as found in Reference \citen{Brennerthesis}. Finally, we note that in Equations~\ref{DISET_current_S} and \ref{Stoof_formula_S}, $t_\mathrm{S}$ and $t_\mathrm{D}$ are indistinguishable, therefore we can only extract their mean contribution when fitting such formulas to experimental data.

\section{Transport characteristics and SET Tuning}\label{supp_D}

\begin{figure}[h!]
\centering
\includegraphics[width=6.2in]{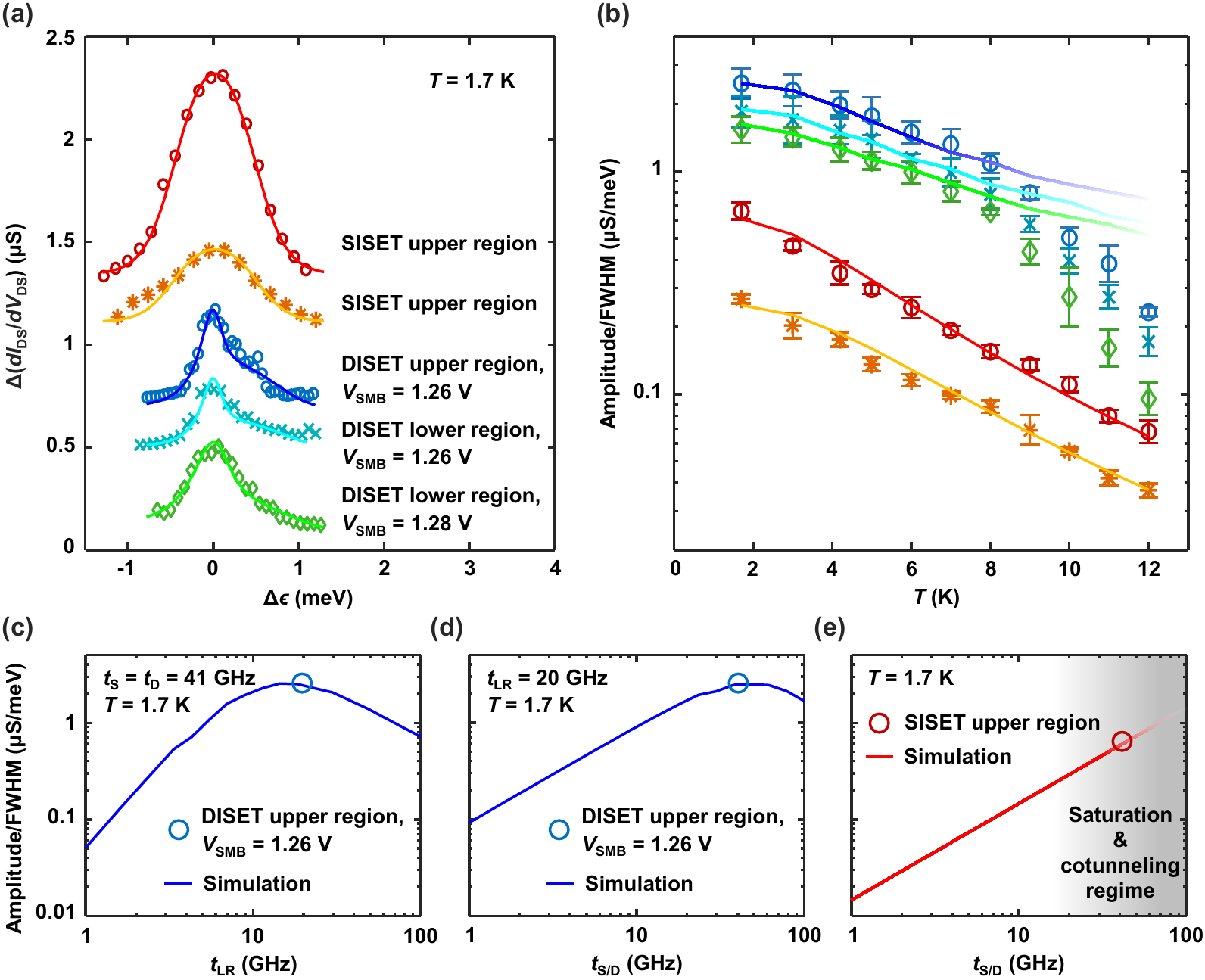}
\caption{Barrier gate dependence of transport characteristics. (a) Measured Coulomb peaks for the SISET and the DISET (markers) with different barrier gate voltages at 1.7\,K. The data are fitted to simulated curves (lines) with corresponding barrier transport rates being the fit parameters. The differential conductance is plotted as a function of $\Delta\epsilon$. (b) Ratio of the peak amplitude to the FWHM, which indicates the peaks’ steepness. The five curves correspond to the five bias configurations in (a). The experimental data (markers) are fitted with simulated curves (lines). (c) Simulated amplitude/FWHM for the DISET as a function of $t_\mathrm{LR}$, with $t_\mathrm{S}=t_\mathrm{D}=41$\,GHz and at $T = 1.7$\,K. The overlaid data point is calculated from the measurement in the DISET‘s upper region, with $V_\mathrm{SMB}=1.26$\,V. (d) Simulated amplitude/FWHM for the DISET as a function of $t_\mathrm{S}$ or $t_\mathrm{D}$, with $t_\mathrm{LR} = 20$\,GHz and at $T = 1.7$\,K. The overlaid data point is calculated from the measurement in the DISET‘s upper region, with $V_\mathrm{SMB}=1.26$\,V. (e) Simulated amplitude/FWHM for the SISET as a function of $t_\mathrm{S}$ or $t_\mathrm{D}$, at $T = 1.7$\,K. The overlaid data point is calculated from the measurement in the SISET‘s upper region.}
\label{figS3}
\end{figure}

    In this section, we elaborate on how the barrier gate voltages influence the transport current in the DISET and the SISET, and explain the dependence using SET transport theory (Equation~\ref{DISET_current} and Equation~\ref{SISET_current} in the main text). This provides useful information for tuning the sensor, especially when it operates as a DISET. In all simulations, we assume a DC source-drain bias of $(1\pm0.2)$\,mV for the DISET and $(0.5\pm0.2)$\,mV for the SISET. The curve fitting are performed in terms of differential conductance. Figure~\ref{figS3}(a) shows a typical peak at each biasing condition at 1.7\,K for both SETs, and Figure~\ref{figS3}(b) shows the temperature dependence of amplitude/FWHM averaged over five peaks at each condition. We estimate the tunnel rates from the fitting of the temperature dependence curve in Figure~\ref{figS3}(b), and estimate the inter-island relaxation rate $\Gamma_\mathrm{LR}$ at a certain temperature from the fitting of the corresponding 1D peaks.

With $V_\mathrm{ST}$ fixed at 2.92\,V, the tunnel rates $t_\mathrm{S}$, $t_\mathrm{LR}$ and $t_\mathrm{D}$ are predominantly determined by $V_\mathrm{SLB}$, $V_\mathrm{SMB}$ and $V_\mathrm{SRB}$. We first study a bias triangle pair in a lower region of the DISET occupancy map, where $V_\mathrm{SLB}$ and $V_\mathrm{SRB}$ are around 1.57\,V and 1.47\,V, respectively. We increase $V_\mathrm{SMB}$ from 1.26\,V to 1.28\,V and observe the corresponding change in a Coulomb peak [Figure~\ref{figS3}(a), cyan and green]. By fitting the curves to Equation~\ref{DISET_current} in the main text, we find that $t_\mathrm{LR}$ increases from $\sim20$\,GHz to $\sim36$\,GHz, while the mean of $t_\mathrm{S}$ and $t_\mathrm{D}$ also increases slightly from $\sim25$\,GHz to $\sim29$\,GHz. This indicates a transition from the regime where inter-island tunneling limits the current to a regime where reservoir-island tunneling is the bottleneck for current. This attests to the control that SMB offers over $t_\mathrm{LR}$, in good agreement with the result in Figure~\ref{figS1}. As can be seen in Figure~\ref{figS3}(a), a higher $V_\mathrm{SMB}$ and resulting higher $t_\mathrm{LR}$ increase both the amplitude and FWHM of the peak.

Additionally, we find that for $V_\mathrm{SMB}=1.26$\,V and 1.28\,V, at 1.7\,K and in the lower occupancy regime, the inter-island relaxation rate $\Gamma_\mathrm{LR}$ are both around 3\,GHz. When the double island is occupied by more electrons, however, $\Gamma_\mathrm{LR}$ increases to $\sim6$\,GHz. The high $\Gamma_\mathrm{LR}$ and the weak dependence on the inter-island coupling can be explained by the small distance between the islands, according to Reference~\citen{Wang2013}. This implies that a higher $\Gamma_\mathrm{LR}$ is primarily related to the increase in island size, which also agrees with our results since the islands become larger in the higher occupation regime.

We study the role of $t_\mathrm{S}$ and $t_\mathrm{D}$ by scanning for peaks in different regions of the DISET occupancy map, while keeping $V_\mathrm{SMB}$ unchanged. In Figure~\ref{figS3}(a), the data in cyan and blue show the peaks measured with $V_\mathrm{SRB}$ around 1.47\,V and 1.57\,V, corresponding to a lower region and an upper region in the occupancy map. $V_\mathrm{SLB}$ remains around 1.58\,V and $V_\mathrm{SMB}$ is fixed at 1.26\,V. As suggested by the curve fitting, the mean of $t_\mathrm{S}$ and $t_\mathrm{D}$ increases from $\sim25$\,GHz to $\sim41$\,GHz, whereas $t_\mathrm{LR}$ increases slightly to $\sim21$\,GHz. Consequently, the peak changes almost only in its amplitude. However, we note that this is valid only when the reservoir-island tunnel coupling is much smaller than the energy scale of $\Delta\epsilon$, otherwise the FWHM will increase significantly due to cotunneling and lifetime broadening.

\begin{figure}[h!]
\centering
\includegraphics[width=4.33in]{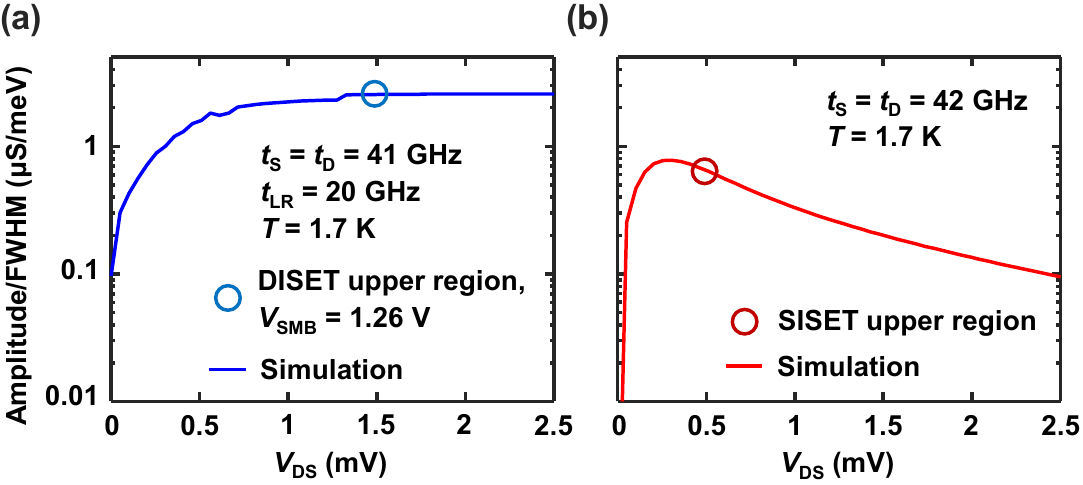}
\caption{Source-drain bias dependence of transport characteristics. (c) Simulated amplitude/FWHM for the DISET as a function of $V_\mathrm{DS}$, with $t_\mathrm{S} = t_\mathrm{D} = 41$\,GHz, $t_\mathrm{LR} = 20$\,GHz and at $T = 1.7$\,K. The overlaid data point is calculated from the measurement in the DISET‘s upper region, with $V_\mathrm{SMB} = 1.26$\,V. (d) Simulated amplitude/FWHM for the SISET as a function of $V_\mathrm{DS}$, with $t_\mathrm{S} = t_\mathrm{D} = 42$\,GHz and at $T = 1.7$\,K. The overlaid data point is calculated from the measurement in the SISET‘s upper region.}
\label{figS4}
\end{figure}

We use our analytical model of the DISET to qualitatively guide the tuning of the DISET to an optimal configuration. We find from Equation~\ref{DISET_current_S} that the peak amplitude in a DISET can be improved by minimizing cotunneling or lifetime broadening by allowing higher $t_\mathrm{S}$ and $t_\mathrm{D}$, provided that the resulting tunnel coupling is much smaller than the energy scale of inter-island detuning $\Delta\epsilon$. Limiting $t_{\mathrm{LR}}$ reduces the peak amplitude, but also limits the lifetime broadening of the peak, which is desirable. An experimental comparison is made in Figure~\ref{figS4} with support from theory. In Figure~\ref{fig2} in the main text, we bias the DISET at the highest $V_\mathrm{SLB}$ and $V_\mathrm{SRB}$ without significant cotunneling or lifetime broadening. We primarily explore different $V_\mathrm{SRB}$ as SRB has the larger lever arm (Supporting Information~\ref{supp_B}), providing a more tunable right island. The DISET peak shown in Figure~\ref{fig2} is taken at $V_\mathrm{SLB}\approx1.58$\,V and $V_\mathrm{SRB}\approx1.57$\,V. The inter-island tunneling $t_\mathrm{LR}$ is lowered by tuning down $V_\mathrm{SMB}$ to the smallest value just before the peak becomes too small or narrow to be accurately measurable. We experimentally find $V_\mathrm{SMB}=1.26$\,V to be a near-optimal value.

In order to transform the two-island arrangement into a single island and operate it as a SISET, we bias the tunneling barriers at $V_{\mathrm{SLB}}=1.54$\,V and $V_{\mathrm{SMB}}=1.26$\,V, while setting $V_{\mathrm{SRB}}=2$\,V to extend the reservoir to the right island location. The dependence on $t_\mathrm{S}$ and $t_\mathrm{D}$ is studied by sweeping ST in different voltage ranges. We measure the Coulomb peak at $V_\mathrm{ST}\approx2.5$\,V and $V_\mathrm{ST}\approx2.8$\,V and plot the data in orange and red in Figure~\ref{figS3}(a). We extract the mean of $t_\mathrm{S}$ and $t_\mathrm{D}$ to be $\sim20$\,GHz for $V_\mathrm{ST}=2.5$\,V and $\sim42$\,GHz for $V_\mathrm{ST}=2.8$\,V. An increase in the reservoir-island tunnel rates results in a larger peak amplitude, similar to the dependence in the DISET.

The dependence of amplitude/FWHM on different tunnel rates is simulated using Equation~\ref{DISET_current} and Equation~\ref{SISET_current} in the main text over a wide range, and plotted in Figure~\ref{figS3}(c)-(e). In (c), we assume that both $t_\mathrm{S}$ and $t_\mathrm{D}$ are 41\,GHz and $T=1.7$\,K, the same conditions under which the peak in the DISET upper region is measured. Sweeping $t_\mathrm{LR}$ from 1\,GHz to 100\,GHz, we see that amplitude/FWHM peaks at $t_\mathrm{LR}\approx15$\,GHz. Similarly in (d), we vary $t_\mathrm{S}$ and $t_\mathrm{D}$ equally and simultaneously, and see a peak at $t_\mathrm{S}=t_\mathrm{D}\approx44$\,GHz. In (e), we study the SISET dependence on $t_\mathrm{S}$ and $t_\mathrm{D}$. Although amplitude/FWHM is theoretically linearly increasing, the data point is taken from the experimentally optimal regime. If $V_\mathrm{ST}$ is further increased, the amplitude becomes saturated and eventually decreases due to strong cotunneling or even saturation. By overlaying the experimental data with the simulated curves, we also confirm that our SET is well-tuned in the comparison of temperature dependence and in the subsequent charge sensing experiment.

We see that the Amplitude to FWHM ratio of a SISET, as seen in Figure~\ref{figS3}(e), improves with a more transparent barrier between island and the leads, while Figure~\ref{figS3}(c) and (d) reveal that the DISET has a specific set optimal tunnel coupling to the leads. This adds an additional constraint in tuning the DISET, but it can be understood and predicted from theory. Note that this difference only plays against DISETs when compared to SISETs, so it does not impact our conclusion that DISET offers a better range of improved readout fidelity compared to SISETs.

Lastly, the dependence of amplitude/FWHM on different source-drain bias $V_\mathrm{DS}$ is also simulated and plotted in Figure~\ref{figS4}. We use the tunnel rates corresponding to the upper region of the DISET and the SISET, which are near-optimal. In Figure~\ref{figS4}(a), we see that amplitude/FWHM of the DISET begins to saturate after $V_\mathrm{DS}\approx0.5$\,mV. This agrees with our experimental observation that the increase in steepness is indistinct when $V_\mathrm{DS}$ is increased from 1\,mV to 1.5\,mV. Further increasing $V_\mathrm{DS}$ only creates larger bias triangles, which is beneficial for finding a sensitive point in charge sensing. Conversely, amplitude/FWHM of the SISET reaches a maximum at $V_\mathrm{DS}\approx0.35$\,mV, followed by a slow decay. In the experiment, we applied a $V_\mathrm{DS}$ of $(0.5\pm0.2)$\,mV, in close proximity to the optimal value.

\section{Output Noise Spectrum}\label{supp_E}

\begin{figure}[h!]
\centering
\includegraphics[width=2.7in]{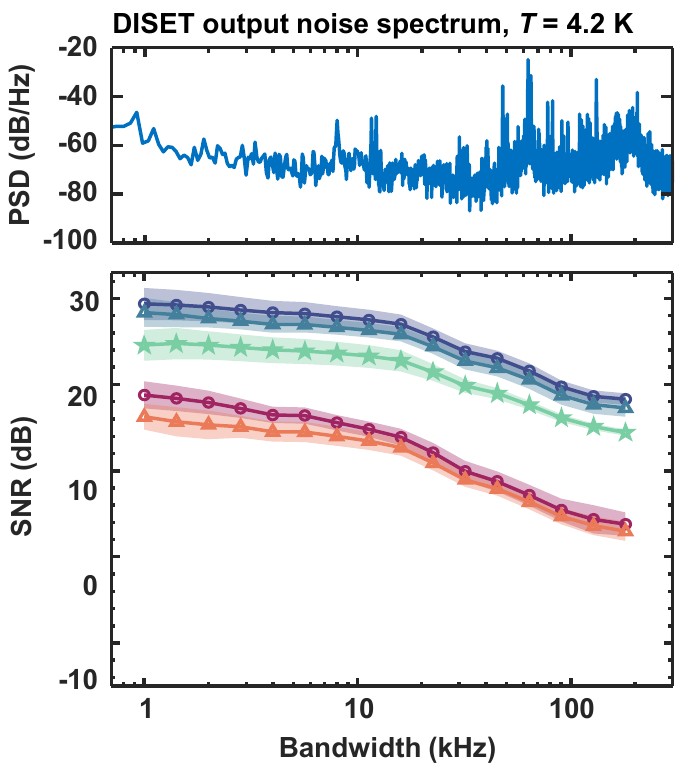}
\caption{Welch power spectral density (PSD) in the single-shot measurement showing the distribution of output noise. The PSD is calculated from a raw current trace without any charge transitions, measured at 4.2\,K with the DISET.}
\label{figS5}
\end{figure}

The SNRs of both SETs experience a rapid decrease after $\sim20$\,kHz, which results in a faster increase of the readout error, especially for the DISET (which is more sensitive to all electric charge shifts, including noise). This can be explained by the noise spectrum shown in Figure~\ref{figS5}. We plot the Welch power spectral density, which represents the noise distribution at different frequencies, on the same frequency scale as the plot of the SNR versus bandwidth. We see an increase in high frequency noise above $\sim20$\,kHz, which originates from the noise in the control setup or other environmental noise.

\twocolumngrid
\newpage
\bibliography{jhuang}

\end{document}